# If I Hear You Correctly: Building and Evaluating Interview Chatbots with Active Listening Skills

**Ziang Xiao[1*], Michelle X. Zhou[2], Wenxi Chen[3], Huahai Yang[3], Changyan Chi[3**]**

[1]zxiao5@illinois.edu, University of Illinois at Urbana-Champaign, Urbana IL, USA
[2]mzhou@acm.org, Juji, Inc., San Jose CA, USA
[3]{wchen, hyang, tchi}@juji-inc.com, Juji, Inc., San Jose CA, USA

## ABSTRACT
Interview chatbots engage users in a text-based conversation to draw out their views and opinions. It is, however, challenging to build effective interview chatbots that can handle user free-text responses to open-ended questions and deliver engaging user experience. As the first step, we are investigating the feasibility and effectiveness of using publicly available, practical AI technologies to build effective interview chatbots. To demonstrate feasibility, we built a prototype scoped to enable interview chatbots with a subset of *active listening* skills—the abilities to comprehend a user's input and respond properly. To evaluate the effectiveness of our prototype, we compared the performance of interview chatbots with or without active listening skills on four common interview topics in a live evaluation with 206 users. Our work presents practical design implications for building effective interview chatbots, hybrid chatbot platforms, and empathetic chatbots beyond interview tasks.

## CSS Concepts
• **Computing methodologies~Intelligent agents**; • **Human-centered computing~Human computer interaction (HCI)**;

## 1 INTRODUCTION
During the past few years, chatbots, which engage users in a one-on-one, text-based conversation, have been adopted for a wide variety of applications [8, 13, 21, 26]. Among various chatbot applications, a promising one is information elicitation (e.g., [51, 63, 64, 69]). For example, Tallyn et al. use a chatbot to elicit user input in an ethnographic study [51]. Li et al. build a chatbot to interview job candidates and aid in talent selection [32]. Recent studies also show several benefits of chatbots for information elicitation, such as eliciting higher quality information than using traditional form-based methods (e.g., [29, 64]).

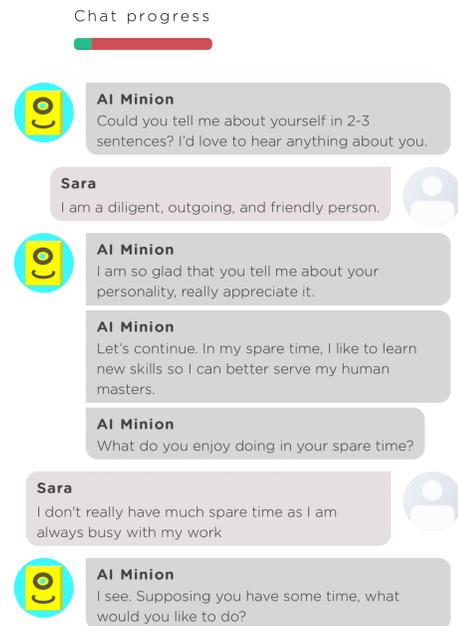

**Figure 1. A screenshot of an example interview conducted by a chatbot (AI Minion) and a user (Sara).**

Inspired by these efforts, we are building interview chatbots to conduct user interviews and facilitate user research. To conduct effective interviews, interview chatbots should have skills similar to that of effective human interviewers [32, 41]. One of such important skills is *active listening*—the abilities to understand and respond to a conversation partner properly [19, 45]. Active listening is shown to facilitate interviews, e.g., eliciting higher quality responses [19, 35, 45] and making an interviewer more socially attractive [59]. In addition, studies find that active listening helps not only oral communication, but also online text communication, including text messaging [2, 3]. Inspired by those findings, we hypothesize that interview chatbots with active listening would be more effective at conducting interviews and engaging interviewees. Figure 1 shows an example of such a chatbot, which can understand the user's input and summarize it in its response, making the user feel heard.

Despite recent advances in Artificial Intelligence (AI), it is still challenging to build capable chatbots [22], let alone create chatbots with active listening skills. Below we highlight three main challenges specific to building effective interview chatbots with active listening skills.

First, it is challenging to build interview chatbots that can effectively grasp and respond to user input to open-ended interview questions, which is the core of active listening. For example, in one of our user surveys, a chatbot asked an open-ended question "*what's the top challenge you're facing*". One user responded:

> "The biggest challenge I've faced is finding a since of purpose. Being around like minded individuals who are constantly wanting more out of life through countless jobs I've never found something I was proud of…"

Another user answered the same question very differently:

> "With a new baby I have a lot of additional expenses. So I have to try to obtain additional income. I try to earn extra income by working on mturk, but the pay is low and I don't like the additional time taken away from my…"

Given such user input, an effective chatbot should respond to each user empathetically to make them feel heard. Few chatbot platforms, however, enable chatbots to handle such complex and diverse user input. For example, popular chatbot platforms like Chatfuel [8] and Manychat [37] hardly handle user free-text input. More advanced platforms like Google Dialogflow [13] and IBM Watson Assistant [25] support Natural Language Processing (NLP), but they often require that a chatbot designer enumerate all user intents to be handled. With such a method, it would be very challenging to build an interview chatbot, since it is difficult to anticipate diverse user responses to open-ended questions and enumerate all possible user intents.

Second, it is difficult to build interview chatbots that can effectively handle complex conversation situations to complete an interview task. As indicated by a recent report, natural language conversations are nonlinear and often go back and forth [22]. In an interview, a user may digress from a planned agenda for various reasons. For example, some users may not understand an interview question and want clarifications (e.g., "*What do you mean*"), while others might dodge a question by responding with "*Why do you want to know*?" or "*I don't know*." Users might also misunderstand a question or simply do not know how to answer it. For example, one user offered an ambiguous response to the question mentioned above:

> "Most challenges are met as an opportunity to grow. Hardest part is losing friends."

Users may also be "uncooperative" and intentionally provide gibberish or irrelevant responses, such as those observed in crowd-sourced user studies [16].

To complete an interview task, a chatbot must "remember" and stick to an interview agenda no matter how many times or how far a conversation has digressed from the agenda. However, most chatbots support scripted dialog trees instead of dynamic, graph-like conversations required by effective interview chatbots.

Third, it is difficult for chatbot designers to take advantage of AI advances due to a lack of AI expertise or resources. For example, deep learning has enabled powerful conversational AI [36, 62, 65, 66, 1] and might help address the first challenge mentioned above. However, these models require large amounts of training data (i.e., interview data), which are hard to acquire.

Given the three challenges mentioned above, we explore new ways to build effective interview chatbots. As the first step, we are investigating the feasibility and effectiveness of using existing AI technologies to build effective interview chatbots with active listening skills.

Our investigation aims at answering two research questions:

**RQ1:** Whether and how can we employ publicly available AI technologies to build effective interview chatbots with active listening skills?

**RQ2**: How effective can such interview chatbots be at handling complex and diverse user input and affecting user experience and interview quality?

To answer the above questions, we have developed a prototype system for building chatbots with active listening skills. We used our prototype to create two chatbots with and without active listening skills, respectively. We evaluated both chatbots live with 206 participants from Amazon Mechanical Turk to compare their performance by a set of metrics, including quality of user responses and user perception.

As will be seen, our answers to the two research questions demonstrate the feasibility and effectiveness of using publicly available AI technologies to build effective interview chatbots. As a result, our work offers three contributions:

1. *Practical approaches to effective interview chatbots.* Our work presents practical implementations to power chatbots with a specific set of active listening skills.

2. A *hybrid framework for developing progressive chatbot platforms.* Our work demonstrates a hybrid chatbot design framework, where rules can be used to bootstrap a chatbot and data-driven models can then be used to improve the chatbot's conversation capabilities.

3. *Design implications for building empathetic chatbots beyond interview tasks.* Since active listening aids effective communications beyond interviews, our work presents design considerations of building empathetic chatbots for a wide variety of applications, including counseling and training, beyond interview tasks.

## 2 RELATED WORK
Our work is related to research in four areas listed below.

### 2.1 Conversational Agents for Information Elicitation
There is a rich line of work on developing conversational agents for information elicitation. These agents roughly fall into three categories: survey bots, AI-powered chatbots, and embodied AI interviewers.

Survey bots text users a set of choice-based questions with little natural language interaction [29, 49, 51]. They normally do not ask open-ended questions nor handle user digressions that may arise during a natural language conversation. More recently, AI-powered chatbots have been used to ask users open-ended questions via texting [20, 30, 63, 64, 69]. Additionally, there is a rich body of work on embodied AI interviewers (e.g., [4, 10, 18, 32, 34, 41, 53, 57]). These AI agents have a human-like form and use both verbal and non-verbal expressions to communicate with users.

Among the three types of conversational agents, our work is most related to AI-powered chatbots. Similar to these efforts, our goal is to deliver engaging interview experience and elicit quality information. However, existing works are limited at handling complex and highly diverse user input [64, 69]. In contrast, our work reported here is intended to improve the conversation capabilities of these chatbots.

### 2.2 Task-Oriented vs. Social Conversational Agents
Although conversational agents have been used in a wide variety of applications, they fall into two broad categories [52]. One type helps users accomplish specific tasks, such as meeting scheduling [22] and information search [69]. The other is to socialize with users without a task (e.g., [35, 62]). Because of the constrained domains and the need for gathering accurate parameters (e.g., meeting time), rule-based approaches are often used to create task-oriented agents [69]. Although a recent data-driven approach to task-oriented agents shows early promises [6], it is not ready for real use. In contrast, data-driven approaches are mostly used to support open-domain, social dialogues [35, 62].

Recently, researchers have developed conversation agents that support both task-oriented and social dialogues in one system [42, 66]. Similar to this line of work, our interview chatbots must support both task-oriented and social conversations during an interview. Unlike this line work, which helps users achieve a task like making a restaurant reservation, interview chatbots must complete its *own* information elicitation task. Such differences impose new challenges on building interview chatbots, such as handling uncooperative users or irrelevant user responses to open-ended questions.

### 2.3 Recent Advances in Conversational Agents
There are numerous computational approaches to building conversational agents, including both symbolic and data-driven approaches [52]. To cope with highly diverse user input, data-driven approaches have been used extensively to handle open-domain conversations. A number of data-driven approaches are used to train retrieval models that find the most probable machine response from a repository of pre-defined responses for a given user input (e.g., [36, 62, 65, 66, 1].) Additionally, generative approaches have been explored to synthesize machine responses that do not exist before (e.g., [46, 47, 65]). However, the quality of generated responses may be erroneous or incoherent, not yet ready for practical applications.

Neither retrieval-based nor generative models alone are practical for building interview chatbots, since they require large amounts of training data—often millions or billions of conversation exchanges [65, 66, 1]. It is difficult to obtain interview data let alone large amounts due to the private or sensitive nature of many interviews. Moreover, a lack of interpretability and control of data-driven results would put an interview chatbot at risk especially in high-stakes contexts, such as customer interviews [54].

To improve interpretability, recently, researchers have explored hybrid approaches. For example, Hu et al. propose to incorporate rules as the weights of neural networks to improve interpretability and performance [24]. Sundararajan et al. propose an approach to identify which input features contribute to the prediction of a deep network [50]. Their approach can extract rules from the networks to help interpret the prediction results and debug the networks. These hybrid approaches have inspired us in developing our prototype, which is perhaps the first of exploring a hybrid framework for building interview chatbots.

### 2.4 Chatbot Platforms
During the past few years, a number of chatbot platforms have been developed to facilitate the creation of chatbots. There are two types of platforms. The first type, including Chatfuel [8] and Manychat [37], supports do-it-yourself chatbot making. However, they have little AI/NLP capabilities and cannot support the creation of interview chatbots with active listening skills. The second type, including Google Dialogflow [13] and IBM Watson Assistant [25], offers AI/NLP capabilities but has a steep learning curve for non-AI experts to use the tools (e.g., they must understand NLP elements such as intents and entities). Moreover, most platforms in this category are designed for making task-oriented chatbots (e.g., restaurant reservation). They must be extended to support interview chatbots to perform tasks and be social at the same time.

Given the limitations of existing chatbot platforms, we decided to extend Juji [28], a chatbot platform that supports both tasks-oriented and social dialogues and allows easy extensions, to build effective interview chatbots. Our decision to extend Juji is detailed in Section 4.3.

## 3 PROTOTYPE OVERVIEW
To demonstrate the feasibility of building effective interview chatbots with existing AI technologies, we have investigated a number of approaches to conversational agents, including both rule-based and data-driven approaches [52].

On the one hand, a rule-based approach uses explicitly coded knowledge (e.g., grammars) to handle user input [38, 60]. Rules are easy to understand and can be used to bootstrap a chatbot. However, it requires expertise and manual efforts to code the knowledge. It is also difficult to code complex, implicit knowledge expressed by users.

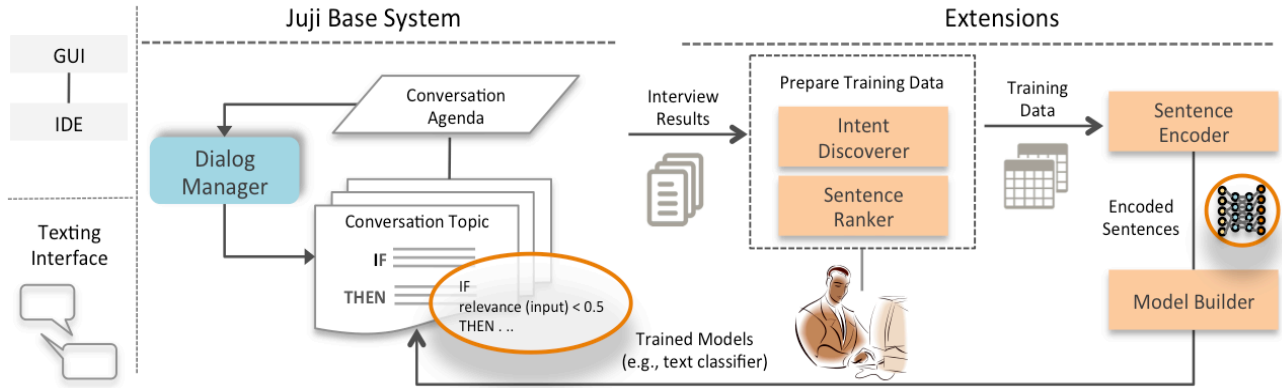

**Figure 2. Overview of our prototype system for building an effective interview chatbot.**

On the other hand, data-driven approaches, such as deep learning, have shown their promises for handling complex and diverse conversations (e.g., [36, 61, 1]). Such approaches, however, require intensive computational resources and large training datasets. Moreover, chatbot designers who are not AI-experts may not know how to fine-tune the approaches or control the use of the results to prevent inappropriate chatbot behavior.

Based on our investigations, a hybrid system that combines rule-based and data-driven approaches seems most promising. Our prototype thus includes two parts: (1) a rule-based system and (2) extensions in support of data-driven models.

As shown in Figure 2, Juji, a publicly available chatbot platform (juji.io), serves as the rule-based system. On Juji, chatbot designers can create, customize, and deploy a chatbot with a graphical user interface (GUI) or an interactive development environment (IDE) [28]. Authoring a Juji chatbot is to define a *conversation agenda* (interview agenda) that contains one or more conversation topics (interview topics) with their temporal order. To drive a conversation, the *dialog manager* activates and manages each topic on an agenda by their temporal order [69]. End users (e.g., interviewees) can interact with a Juji chatbot via a texting interface on a website or via Facebook Messenger.

Our prototype also includes a set of extensions that enable the incorporation of data-driven approaches from two aspects: (a) preparing training data and (b) training models to handle user free-text input. To prepare training data, the *intent discoverer* and the *sentence ranker* work together to *automatically* identify user intents and label training data. Since automated analyses are often imperfect, a human can examine and rectify mislabeled data.

The *sentence encoder* encodes labeled training data into fixed-length dense vectors to capture rich linguistic features. The *model builder* then uses the encoded data to train various models, such as text classification models for user intent prediction. The trained models can then be used to control chatbot behavior. For example, one model may predict how semantically relevant a user input is to an open-ended interview question. Based on the prediction, a chatbot can respond properly during an interview.

## 4 PROTOTYPE DESIGN AND KEY COMPONENTS

Here we present the scope and technical focus of our prototype along with our design criteria. We then describe the key components of our prototype.

### 4.1 Prototype Scope and Technical Focus

To enable active listening, our technical focus is on interpreting the semantics of user free-text responses. Active listening can be carried out through a number of communication techniques, such as paraphrasing, verbalizing emotions, and summarizing [1, 9]. No matter which technique is used, an interview chatbot must comprehend the semantics of a user response before it can respond to it properly. To *encourage* a user to provide a better response, for example, a chatbot must at least understand how semantically relevant the user response is to its question. Similarly, a chatbot must understand the underlying semantics of a user response to *summarize* the response.

As a start, we focus on a subset of the techniques (Table 1). For example, we exclude impromptu techniques, such as *balancing* [1], since currently we build chatbots for only structured interviewing with pre-defined questions.

| Technique | Synopsis | Example |
|---|---|---|
| Paraphrasing | Restate a user input to convey understanding. | *"I see you love to hang out with your friends."* |
| Verbalizing Emotions | Reflect a user's emotions in words to show empathy. | *"I can tell intellectual activities make you happy. Just keep doing what you love."* |
| Summarizing | Summarize the key ideas stated by a user to convey understanding. | *"If I hear you right, you care about others and have great leadership potential."* |
| Encouraging | Offer ideas and suggestions to encourage conversation. | *"You've made an interesting point, could you elaborate?"* |

**Table 1. Active listening supported by our prototype.**

### 4.2 Prototype Design Criteria

When designing our prototype, we faced many choices. In general, our decisions were guided by three criteria:

(1) *Reproducibility*. We select only publicly available technologies so other researchers and practitioners can easily reproduce our work.

(2) *Practicality*. We prefer technologies that require low resources and little training or customization.

(3) *Adoption*. We favor technologies that non-AI experts can use to build custom interview chatbots.

### 4.3 The Juji Base System

Among many chatbot platforms, we chose to extend Juji for several reasons. First, Juji meets all three design criteria described above. Specifically, Juji is publicly available, and its rule-based system allows a chatbot designer to bootstrap a chatbot without training data. Second, Juji has demonstrated its success in support of interview chatbots. Several recent studies show various interview applications of Juji chatbots, such as job candidate interview [32], market research interview [64], and student interview [20, 32, 63]. In addition, we chose Juji for another two important reasons.

*4.3.1 Ability to Manage Nonlinear Conversations*

Conducting interviews is a complex task, which must handle a task-oriented (e.g., [69]) *and* an open-domain social dialog (e.g., [36, 62]). Specifically, an interview chatbot has a task (interview) to complete. An interview by nature is also a social dialog during which an interviewee may "wander off" from an interview topic. Here we use *side talking* to refer to any conversations outside the pre-defined interview topics on an interview agenda.

To ensure interview quality, an effective interview chatbot must recognize and handle side talking properly, no matter where and how many times it occurs during an interview. Juji is the only chatbot platform that we know and can handle complex conversation flows automatically. Specifically, Juji keeps track of a conversation context automatically and always brings a conversation back on track from side talking [63, 1]. Table 2 is an example showing how Juji handles multiple user interruptions and still manages to bring the user back to its original question (see video).

*4.3.2 Extensibility*

We selected Juji also for its extensibility as its rules take third-party functions or API [69]. For example, an external gibberish-detection function can be embedded in a rule to handle user gibberish input. Moreover, Juji topic-based conversation model allows us to make targeted, topic-specific improvements. Compared to improving end-to-end dialogs (e.g., [36, 62]), supporting topic-specific improvements has two benefits. First, it requires much less training data to achieve good performance (see Section 5.1). Second, it avoids propagations of modifications or errors to other topics. In contrast, in an end-to-end approach, training data may improve certain parts of a dialog but adversely affect other parts. It is often difficult to control such effects.

| | |
|---|---|
| Chatbot: | I'd love to learn what kind of reader you are. What types of books do you like to read? |
| User: | I don't know. What about you? |
| Chatbot: | Sorry, I cannot read yet. Could we go back to my question? |
| User: | What was your question? |
| Chatbot: | I was asking: What types of books do you like to read? |
| User: | It's really hard to say since I read a lot. |
| Chatbot: | No worries, just share what's on your mind. |
| User: | I guess my favorite kind would be sci-fis. |

**Table 2. How Juji handles a nonlinear conversation.**

### 4.4 The Extensions: Data-Driven User Intent Prediction

To power interview chatbots with active listening skills (Table 1), the key focus of our prototype is to interpret the meaning of complex and diverse user input. Since pattern-based rules cannot handle such user input, we have incorporated data-driven approaches to semantic interpretation. However, interpreting fine-grained semantics of natural language expressions is still challenging even with data-driven approaches. We thus scaled down the challenge to focus on identifying the *semantic gist* of user input—the high-level intent that users imply.

Recent advances show the effectiveness of using text classification for identifying implied user intent (e.g., [23, 36, 62, 1]). Additionally, text classification can process text at scale, which is important in our application since an interview chatbot may converse with thousands of users at the same time. Rich public resources are also available for building text classification models (e.g., Google developer resource). Our extensions thus focus on supporting text classification models to auto-identify user intent.

To facilitate the construction of text classification models especially for non-AI experts, we have built a set of components to support a three-step model construction: (a) preparing training data, (b) encoding training data, and (c) training text classification models.

*4.4.1 Preparing Training Data: Machine-Aided Labeling*

A key step for building text classification models is to obtain *labeled* training data. User responses related to an interview question (topic) can serve as the training data for building topic-specific classification models. For example, on the interview topic "*What's your top challenge*", user responses to this question can be used as training data to build classification models for improving the conversation around this topic. However, labeling user responses is non-trivial. First, these responses are highly diverse. It is difficult for humans to consistently code the intent for a large number of responses. Second, user responses may be complex and convey *multiple* intents simultaneously, which makes human coding even harder.

To facilitate data labeling, we support a 3-step process: *auto-labeling* (2 steps) and human validation (last step).

**Step 1. LDA-based Intent Discovery**

To classify user intents, we first identify the intents conveyed by training data. We chose to use Latent Dirichlet Allocation (LDA) model for this task because LDA is effective at extracting hidden intents (topics) [5]. It is an unsupervised technique and requires no training. Moreover, LDA implementations are publicly available.

In our prototype, we implemented LDA with Gensim [42], an open-source library. Given a set of training data (user responses), LDA automatically derives a set of intents (topics) to summarize these responses (documents). Since these extracted intents are unordered, we enhanced the LDA results by ranking the intents by their coverage [48]. For example, the LDA analysis of 2680 user responses to "*what's your top challenge*" identified top-3 intents by coverage: coping with changes (23.15% of user responses), people problems (21.27%), and time management issues (15.03%).

**Step 2. Centroid-based Sentence Ranking**

The LDA-extracted intents (topics) are typically summarized by a set of keywords. However, it is difficult to label the intents by the associated keywords due to missing context [33]. On the other hand, a user response often conveys multiple intents, it is difficult to identify representative user responses (positive examples) of a given intent.

We thus enhance the LDA results to automatically rank user responses by their semantic proximity to an intent. In particular, we implemented a centroid-based approach. Given an identified intent, we first clustered user responses whose probability distribution over the intent exceeds a threshold. We then used LexRank to rank responses based on their lexical centrality and semantic proximity to the cluster centroid [14]. The higher ranked responses can be considered positive examples for identifying the intent. The lower-ranked ones can then be used as negative examples.

**Step 3. Human Validation**

The two steps described above auto-label user responses as positive and negative examples of a user intent. The results however may not always be correct. A human should always validate the auto-labeled data. A human could always adjust the parameters used in steps 1-2 (e.g., adjusting the number of user intents to be identified by LDA) to redo the analysis and obtain new labeled data.

*4.4.2 Encoding Training Data: Sentence Embedding*
To train text classification models, we need to represent the training data uniformly. We experimented with several encoders and chose to use the publicly available Universal Sentence Encoder (USE) in the Google TensorFlow library to encode each example [7]. We made this choice for several reasons. First, the encoder is trained and optimized for representing longer text like ours. It automatically captures rich, latent features in the text. Second, it is trained on a wide variety of data sources and generalized for diverse NLP tasks. Third, the library publishes its internal variables so we can fine-tune the model with our own data.

Moreover, USE meets our design criteria better than other encoders especially by balancing performance and resource requirements. For example, we experimented with the last layer of pre-trained BERT-Large model for sentence embedding [11]. It produced comparative performance (e.g., F1 scores) but required twice the size of the USE to store the smallest encoding. Although we can fine-tune encoders to improve their performance [11], we did not do so in the prototype reported here, since our goal is to examine the effectiveness of the technologies as is—"lowest-hanging fruits" before exploring alternatives (Section 6.5).

*4.4.3 Training Text Classification Models*
Given the encoded training data, training text classification models is straightforward. For their interpretability, performance, and requirement on training data, we trained binary classification models with a probability score [15, 27].

**4.5 Enabling Active Listening**
Since Juji rules take external functions, we incorporate the trained classification models into the rules associated with specific conversation topics. The rules will be triggered by the prediction results at run time to guide the generation of proper system responses, enabling active listening. Below we use a concrete example to demonstrate such use.

Consider the interview topic *what's your top challenge*. Four classification models are trained to process user input and handle four high-level user intents on this topic. The first model (*Relevance*) predicts whether a user response is relevant to the topic, and another three predict whether a user response implies one of three intents, respectively: *time management issue* (C1), *people problems* (C2), and *coping with changes* (C3). To incorporate these models, we add a rule attached to the topic:

| IF | Relevance(?u) > threshold1 && (?model ← (max (C1(?u), C2(?u), C3(?u)) > threshold2)) |
|---|---|
| THEN | generate-response (?model) |

The above rule states that if a user input (?u) is relevant and implies one or more of the three intents, the chatbot generates a response based on the best-detected intent.

Assume a user input:

> "*I think the main challenge will be starting a new job, ... I will have to learn new ways of doing things, and starting over is not always easy.*"

In this case, model C3 "*coping with changes*" would produce the highest probability. The chatbot then generates a set of system response candidates based on the identified semantic intent and active listening techniques (Table 1). Below is a set of responses generated using the *summarizing* technique on the "*coping with changes*" intent:

*"Your description really resonates with me as I also struggle coping with changes or new settings."*

*"I would feel the same in your situation since handling new things is always challenging. Thanks for sharing."*

*"Coping with changes is always hard and I wish I could help you in such situations once I become smarter."*

Currently, a response is randomly selected [32]. Juji always offers a default response if no user intent can be predicted with a certain level of confidence.

## 5 EVALUATION

To evaluate the performance of our prototype, we have conducted extensive experiments. Here we report two sets of results: (1) effectiveness of predicting user intent, and (2) impact on user response and experience.

### 5.1 Evaluating User Intent Prediction Models

We conducted a set of experiments to measure how well text classification models can predict implied user intent. The accuracy of these models directly determines how well an interview chatbot can listen actively during an interview.

We have been using the Juji base system to build chatbots for various real-world applications, such as student surveys. In these applications, we used Juji built-in topics [69] as is. From these applications, we have accumulated a number of user responses on various topics. We decided to test our prototype on improving four most used interview topics:

**Q1:** *Could you tell me about yourself in 2-3 sentences?*

**Q2:** *What do you enjoy doing in your spare time?*

**Q3:** *What is the best thing about you?*

**Q4:** *What is the biggest challenge you face now?*

We chose these four topics for additional reasons. First, they often appear in an interview to build rapport [4] and usually elicit diverse user responses. Handling these topics well can benefit many real-world applications. Second, handling more specific interview topics (e.g., asking about one's work experience) may require deeper domain knowledge and in turn more advanced AI/NLP. Since we did not know how well off-the-shelf AI would work, we focused on improving general interview topics first. Third, our extensions require training data, and these four topics gathered the most training data from real interviews.

For each interview topic, we created a training data set by randomly selecting about 4000 user responses on that topic. Our enhanced LDA model first analyzed its training set and identified 4-5 implied intents, each of which covered at least 10% training data. For each identified intent, the centroid-based analysis produced a set of ranked responses by their semantic proximity to the intent and auto-labeled the top 20% as positive examples and the bottom 20% as negative examples. A human then verified and amended the labels if needed. As a result, a total of 17 user intents were identified across four interview topics, with about 1000 labeled training samples per intent.

We then trained a total of 68 text classification models for 17 intents, each with four popular classification models: logistic regression, linear SVM, Adaboost, and Naïve Bayes. For each of 68 models, we performed stratified 10-fold cross validations and examined four standard performance metrics: *precision*, *recall*, *F1*, and *accuracy*. Since we cannot fit all 68 sets of results in the paper, here we report the overall results and two representative sets.

Table 3 shows the averaged performance of best models per topic. Overall, logistic regression and linear SVM performed the best across all data sets. Logistic regression performed the best over more heterogeneous data sets, while linear SVM performed the best on more homogeneous data sets. Table 4 shows two such example results. At the top of the table, logistic regression performed the best over heterogeneous expressions on leadership—user responses to "*what's your top talent*". These responses are syntactically diverse and semantically complex. Here are two responses:

*"I always think through things and try to do everything I care for the people around me and those I care about. I am quick to give advice and emotional support, and I will always give my time when I feel that I have it"*

*"I am an encouraging person and I know that I can help others and unite a group that is starting with myself. … I will give my all 100 % of the time. I believe not only in myself but in others and I think that is something really impactful when it comes to moving forward..."*

On the bottom of the Table 4, linear SVM performed the best over more homogeneous expressions on *hanging out with friends*—user responses to "*what do you enjoy doing in your spare time*". Here are two examples:

*"I like to spend time with my friends—we talk or do fun outdoor activities together."*

*"I enjoy spending time with friends on the weekends. ... surfing for me was a great way to make friends."*

### 5.2 Live Chatbot Evaluation

We also designed and conducted a between-subject study that compared the live performance of chatbots with and without active listening.

| | Precision | Recall | F1 | Accuracy |
|---|---|---|---|---|
| **Q1 (self intro)** | 0.7661 | 0.8433 | 0.8001 | 0.8506 |
| **Q2 (hobbies)** | 0.9003 | 0.7995 | 0.8332 | 0.9004 |
| **Q3 (best about u)** | 0.8061 | 0.8367 | 0.8199 | 0.8098 |
| **Q4 (top challenge)** | 0.8653 | 0.5728 | 0.6543 | 0.9215 |

Table 3. Stratified 10-fold cross validation for four topics. Logistic Regression for Q1, and SVM for Q2, Q3, and Q4.

### 5.2.1 Study Design

We designed a 10-minute interview on six topics, including the four topics mentioned above. After chatting on each topic, users were asked to rate how well the chatbot understood them. In addition to the four topics, users were asked of their opinion about the chatbot and what the chatbot could do for them. Near the end of an interview, users were asked to rate the chatbot on two more aspects: their interest in chatting with the chatbot in the future and their overall chat experience. All the ratings were on a 5-point Likert scale, 1 being poor and 5 being excellent. We also collected user basic demographics, such as gender and age group.

We built two chatbots. One was built with only the Juji base system and served as the baseline (Baseline). The other (Full Version) was first bootstrapped by the Juji based system and then improved by the text classification models mentioned above. This chatbot demonstrated active listening on the four interview topics. Both chatbots asked the same interview questions in the same order.

### 5.2.2 Participants

We recruited participants on Amazon Mechanical Turk with an approval rate equal to or greater than 99% and located in the U.S. or Canada. We paid each participant $12.5/hr. The participants were given the same message, "interviewing with an AI chatbot", and one of two chatbot URLs that was randomly assigned.

### 5.2.3 Measures

Active listening aims at improving interviewee experience and interview quality. We thus compared the effects of the two chatbots on these aspects by a set of metrics adopted from previous work [35, 64].

*Engagement duration*. It measures how long a user engages with a chatbot, implying a user's willingness to engage and the quality of interaction [35].

*Response length*. This counts the total number of words in a user's text responses, indicating a user's willingness to engage especially in an interview [35].

| **Leadership** (positive examples: 567, negative examples: 479) | | | | |
|---|---|---|---|---|
| | Precision | Recall | F1 | Accuracy |
| **Logistic Regression** | **0.7795** | 0.7654 | 0.7724 | **0.7534** |
| **Linear SVM** | 0.7624 | **0.7865** | **0.7743** | 0.7505 |
| **AdaBoost** | 0.7463 | 0.7462 | 0.7462 | 0.7228 |
| **Naïve Bayes** | 0.7402 | 0.7457 | 0.7429 | 0.7188 |
| **Hangout w/ friends** (positive examples: 529, negative examples: 492) | | | | |
| **Logistic Regression** | 0.9118 | 0.8865 | 0.8990 | 0.8962 |
| **Linear SVM** | **0.9167** | **0.8940** | **0.9052** | **0.9020** |
| **Adaboost** | 0.8593 | 0.8506 | 0.8549 | 0.8502 |
| **Naïve Bayes** | 0.8257 | 0.8657 | 0.8452 | 0.8345 |

Table 4. Stratified 10-fold cross validations for two data sets.

| | Precision | Recall | F1 | Accuracy |
|---|---|---|---|---|
| **Q1 (self intro)** | 0.7673 | 0.6784 | 0.6810 | 0.7993 |
| **Q2 (hobbies)** | 0.8634 | 0.8536 | 0.8372 | 0.9267 |
| **Q3 (best about u)** | 0.6065 | 0.8064 | 0.6246 | 0.8007 |
| **Q4 (top challenge)** | 0.6357 | 0.3605 | 0.4268 | 0.9237 |

Table 5. Intent prediction for four interview topics. Logistic Regression for Q1 and Q3, and SVM for Q2 and Q4. The accuracy metric was biased due to many true negatives.

*Response informativeness*. To estimate the quality of an interview, we computed the *informativeness* of each user's text responses ($R$) to measure the richness of information contained in the responses by "bits" [64].

*Response Quality Index (RQI)*. To measure the response quality of each participant, we created a *Response Quality Index (RQI)* based on [64]. It measures the overall response quality of $N$ responses given by a participant on three dimensions (specificity, relevance, and clarity):

$$RQI = \sum_{i=1}^{N} relevance[i] \times clarity[i] \times specificity[i]$$

*User ratings*. For each participant, we computed three 5-point Likert scale. One rated the comprehension of a chatbot and an index (*agentC*) was created by adding up the user's ratings on each of the four topics. Another (*interestR*) rated a user's interest of chatting with a chatbot, while the third (*chatR*) rated the user's chat experience.

### 5.2.4 Results

We received a total of 206 completed interviews: 108 engaged with the full version of the chatbot (56% female, 44% male), while 98 interacted with the baseline (37% female and 63% male). The participants were all above 18, and 134 (65%) of them were between 18-34 years old. On average, the participants spent 9.33 minutes with the full version of the chatbot and 7.82 minutes with the baseline. To compute *RQI* for each participant, we manually coded each response following a similar process described in [64]. A total of 824 responses were coded, each with three dimensions on a 3-point Likert scale: 0 (bad), 1 (ok), and 2 (good).

To compare the true effect of two chatbots, we chose to use ANCOVA analyses [43]. In each analysis, the independent variable was the version of chatbot used, and the dependent variable was one of the measures described above. Each analysis was controlled for demographics, which may influence the results, according to previous studies [39].

Table 6 summarizes the results. The chatbot with active listening skills (full version) outperformed the baseline *significantly* across all measures. The full version scored higher on agent comprehension (*agentC*), user interest (*interestR*), and user chat experience (*chatR*). It also chatted with the participants longer (*engagement duration*), and elicited more words

|  | Full Version | | Baseline | | $F$ | $p$ | $\eta_p^2$ |
| --- | --- | --- | --- | --- | --- | --- | --- |
|  | $M$ | $SD$ | $M$ | $SD$ |  |  |  |
| *Engagement Duration* (mins) | 9.33 | 4.47 | 7.82 | 4.85 | $F(1,201)=5.79$ | <0.05* | 0.03 |
| *Response Length* (words) | 125.70 | 56.96 | 100.65 | 48.57 | $F(1,201)=11.72$ | <0.01** | 0.05 |
| *Informativeness* (bits) | 109.66 | 52.32 | 90.21 | 43.78 | $F(1,201)=8.29$ | <0.01** | 0.05 |
| *Response Quality Index (RQI)* | 21.28 | 6.00 | 16.83 | 6.28 | $F(1, 201)=26.98$ | <0.01** | 0.12 |
| *Agent Comprehension (agentC)* | 13.22 | 3.74 | 11.13 | 4.28 | $F(1, 202)=13.87$ | <0.01** | 0.06 |
| *User Interest (interestR)* | 3.56 | 1.24 | 3.07 | 1.29 | $F(1, 202)=7.79$ | <0.01** | 0.04 |
| *User Chat Experience (chatR)* | 3.81 | 1.03 | 3.19 | 1.20 | $F(1, 202)=15.97$ | <0.01** | 0.07 |

a. All results were controlled for participant demographics, including gender and age group.
b. Results for *Informativeness*, *RQI*, and *Response Length* were also controlled for *Engagement Duration*.
c. Results for *Engagement Duration* were also controlled for *Response Length.*

**Table 6. Comparison results of two versions of chatbots with ANCOVA analyses.**

(*response length*), richer information (*informativeness*), and higher-quality responses (*RQI*).

Additionally, we evaluated the performance of predicting user intents from the participants' input. Since there is no space to list individual prediction results for all 68 models, Table 5 shows the averaged performance of the best prediction models for each topic. The models performed worse in the real world than the cross validations for Q3 and Q4, while showed comparable results in Q1 and Q2 (Table 3).

Overall, the models used for predicting user intents on *Q2* (hobbies) performed the best, while the models for *Q4* (top-challenge) performed the worst. In *Q4*, the participants' responses were *very different* from the training data, which was collected mostly from student interviews. For example, many participants talked about their financial challenges, which were not covered by our training data. While it is difficult to anticipate user responses to an open-ended interview topic, our approach allows fast incremental improvements. Specifically, our extensions can analyze and label new user responses, and then use them to train classification models for predicting new user intents and improving chatbot capabilities.

### 5.3 Summary of Findings
Our evaluations helped answer our two research questions.

**RQ1:** It is feasible to use existing and practical AI technologies to build effective interview chatbots with active listening skills.

**RQ2:** Chatbots with active listening skills are more effective at engaging users and eliciting quality user responses, compared to those without such skills.

## 6 LIMITATIONS AND FUTURE WORK
While our evaluations are encouraging, they reveal several limitations. Below we discuss these limitations and future work to overcome the limitations.

### 6.1 Effect of Individual Active Listening Techniques
Our work compared the performance of chatbots with or without active listening skills but not on how *each* skill impacts chatbot performance. Consider a user input "*I like outdoor activities, such as hiking and running*". A chatbot can demonstrate active listening in multiple ways, e.g., *verbalizing emotions* (VE) versus *summarizing* the content (S):

*I can tell physical activities make you happy* (VE)

*It seems you are physically active* (S)

Currently, a chatbot randomly selects one. To make better use of different active listening techniques, we need to investigate their individual effect on different users (e.g., emotional vs. calm person) and on different interview topics (e.g., a sensitive topic discussing one's hardship versus a generic topic on one's hobbies).

### 6.2 Interpreting Deeper User Intents
Our prototype supports the interpretation of high-level, hidden user intents—the semantic gist of user input. Several participants in our study voiced that certain chatbot responses were vague and shallow. To generate more meaningful responses, a chatbot must extract deeper user intents, such as the conveyed semantic concepts and relationships among the concepts. Recent work on knowledge embedding that incorporates knowledge graphs with neural networks may offer a potential solution [55].

### 6.3 Interrelating Interview Topics
Based on Juji's topic model, our chatbots treat each interview topic independently. In reality, interview topics may be related. For example, in our study, a participant made a self-introduction as follows:

"*I'm currently a student. I like to watch football in my spare time, and study I guess.*"

Later, the participant was annoyed by the hobby question, since he thought he already answered it during the self-intro. Currently, the self-intro and the hobby are considered two separate topics. It does not use a user's input given in one topic to influence the discussion on another. Nonetheless, this made the participant feel that the chatbot did not pay attention to his input or could not remember it.

To improve this situation, an interview chatbot needs to remember a user's input, and use it to guide follow-on conversations. One potential solution is to build a knowledge graph to relate different interview topics. This would however require that user input be parsed into a knowledge graph that can be retrieved and reasoned [40]. The main challenge is to determine what knowledge entities (e.g., hobby) should be extracted, since user free-text responses given in an interview are complex and highly diverse.

### 6.4 Active Listening by Asking
One key active listening technique not supported by our prototype is to ask impromptu questions based on a user response to deepen a conversation [35]. This requires that a chatbot *automatically* come up with follow-up questions based on a user's input. Although there is research on question generation, it is often done in a static context [1, 12] or without a conversation goal [56]. We are exploring how to generate effective questions in a highly dynamic yet goal-oriented context like interviewing. This will enable an interview chatbot to follow on interesting ideas that emerged during a conversation and discover unexpected, new insights.

### 6.5 Experimenting with Alternatives
When building our prototype, we made careful technical choices by our design criteria (Section 4.2). Nevertheless, we recognize there are many alternatives that may be equally or more effective. For example, instead of using LDA to discover user intents, we could explore correlation explanation that requires less domain knowledge of the data [7]. We could also try fine-tuning sentence encodings [11] or training sentence embedding and text classification jointly [57] to see if we could achieve better performance. As all these explorations require deeper expertise and more resources, we will do so in the near future and compare their effect on the performance of interview chatbots.

### 6.6 Evaluating Prototype Usability
Although we wish to help non-AI experts build effective interview chatbots, we have not yet evaluated the usability of our system for two reasons. First, we want to verify the feasibility and effect of the prototype before evaluating its usability. Second, our current prototype reuses the Juji GUI for customizing and deploying a chatbot, while providing a command-line interface for data labeling and model training. We feel this UI combination is cumbersome and a more integrated UI is needed as we advance our prototype.

### 6.7 Improving Specific Interview Topics
Our study focused on evaluating the enhancements of four commonly used interview topics. Our results also show that our approach performed better on general topics. This means the current approach with the off-the-shelf-AI and small training data may be limited at handling topics involving deeper domain knowledge (e.g., discussing one's work experience). To develop better approaches, we need to examine how our approach performs across a wider range of interview topics. Moreover, evaluating our system in the real world with actual interviewees may further inform us about the generalizability of our approach.

## 7 DESIGN IMPLICATIONS
Our work demonstrates the feasibility of building interview chatbots with active listening skills and the effectiveness of such chatbots. It thus presents several design implications for building better chatbots and chatbot platforms.

### 7.1 Practical Approaches to Effective Interview Chatbots
Our evaluation shows the effectiveness of an interview chatbot with active listening skills better at engaging users and eliciting quality user input. It implies that a hybrid approach as we developed can power interview chatbots with active listening skills. Moreover, data-driven models can be effectively trained for each topic with small training data sets, more practical than training end-to-end dialogs with large data sets. Although we did our implementations on top of Juji, our methodology is platform agonistic since it can be used to extend any chatbot platforms.

### 7.2 Hybrid and Progressive Chatbot Platforms
Our prototype shows how to support incremental chatbot improvements. Specifically, its hybrid chatbot design framework enables designers to bootstrap a chatbot with rules, use it to collect training data, and train models to improve it. Moreover, chatbot platforms should offer utilities like LDA-based intent discovery to help designers label training data and facilitate the improvement cycle.

### 7.3 Building Empathetic Chatbots Beyond Interviewing
Active listening is used widely in situations like counseling and training [19, 45] beyond interviewing. An AI counselor or coach can also be powered with active listening skills to be more effective in their tasks. So far, few systems support the easy creation and customization of empathetic AI agents with active listening skills. Our work is a stepping-stone toward this direction to enable non-AI experts to create effective chatbots with active listening skills for a wide variety of applications beyond interview tasks.

## 8 CONCLUSION
To investigate the feasibility of using publicly available technologies for building effective interview chatbots and the effect of such chatbots, we have presented a prototype that combines a rule-based chatbot builder with data-driven models to power interview chatbots with active listening skills. These skills enable a chatbot to better handle complex and diverse user responses to open-ended interview questions. As a result, such a chatbot delivers more engaging user experience and elicit higher-quality user responses.

## ACKNOWLEDGMENT
This work is in part supported by the Air Force Office of Basic Research under FA9550-15-C-0032.

# REFERENCES

[1]. Jun Araki, Dheeraj Rajagopal, Sreecharan Sankaranarayanan, Susan Holm, Yukari Yamakawa, and Teruko Mitamura. 2016. Generating questions and multiple-choice answers using semantic analysis of texts. In *Proceedings of COLING 2016, the 26th International Conference on Computational Linguistics: Technical Papers*. 1125–1136.

[2]. Christine Bauer, Kathrin Figl, and Renate Motschnig-Pitrik. 2010. Introducing'Active Listening'to Instant Messaging and E-mail: Benefits and Limitations. *IADIS International Journal on WWW/Internet* 7, 2 (2010), 1–17.

[3]. Christine Bauer and Kathrin Figl. 2008. Active listening" in written online communication-a case study in a course on "soft skills for computer scientists. In *2008 38th Annual Frontiers in Education Conference*. IEEE, F2C–1.

[4]. Timothy Bickmore. 2010. Relational agents for chronic disease self management. *Health Informatics: A Patient-Centered Approach to Diabetes* (2010), 181–204.

[5]. David M Blei, Andrew Y Ng, and Michael I Jordan. 2003. Latent dirichlet allocation. *Journal of machine Learning research* 3, Jan (2003), 993–1022.

[6]. Antoine Bordes, Y-Lan Boureau, and Jason Weston, 2017. Learning end-to-end goal-oriented dialog. *ICLR*' 2017.

[7]. Daniel Cer, Yinfei Yang, Sheng-yi Kong, Nan Hua, Nicole Limtiaco, Rhomni St John, Noah Constant, Mario Guajardo-Cespedes, Steve Yuan, Chris Tar, and others. 2018. Universal sentence encoder. *arXiv preprint arXiv:1803.11175* (2018).

[8]. Chatfuel. 2019. Retrieved from https://chatfuel.com/

[9]. Bert Decker., 1989. How to communicate effectively, Page, London, UK.

[10]. David DeVault, Ron Artstein, Grace Benn, Teresa Dey, Ed Fast, Alesia Gainer, Kallirroi Georgila, Jon Gratch, Arno Hartholt, Margaux Lhommet, and others. 2014. SimSensei Kiosk: A virtual human interviewer for healthcare decision support. In *Proceedings of the 2014 international conference on Autonomous agents and multi-agent systems*. International Foundation for Autonomous Agents and Multiagent Systems, 1061–1068.

[11]. Jacob Devlin, Ming-Wei Chang, Kenton Lee, and Kristina Toutanova. 2019. BERT: Pre-training of Deep Bidirectional Transformers for Language Understanding. In *Proceedings of the 2019 Conference of the North American Chapter of the Association for Computational Linguistics: Human Language Technologies, Volume 1 (Long and Short Papers)*. 4171–4186.

[12]. Xinya Du, Junru Shao, and Claire Cardie. 2017. Learning to Ask: Neural Question Generation for Reading Comprehension. In *Proceedings of the 55th Annual Meeting of the Association for Computational Linguistics (Volume 1: Long Papers)*. 1342–1352.

[13]. Dialogflow. 2019. Retrieved from https://dialogflow.com/

[14]. Günes Erkan and Dragomir R Radev. 2004. Lexrank: Graph-based lexical centrality as salience in text summarization. *Journal of artificial intelligence research* 22 (2004), 457–479.

[15]. Jerome Friedman, Trevor Hastie, and Robert Tibshirani, 2001. *The Elements of Statistical Learning* (Vol. 1, No. 10). New York, NY, USA:: Springer series in statistics.

[16]. Ujwal Gadiraju, Ricardo Kawase, Stefan Dietze, and Gianluca Demartini. 2015. Understanding malicious behavior in crowdsourcing platforms: The case of online surveys. In *Proceedings of the 33rd Annual ACM Conference on Human Factors in Computing Systems*. ACM, 1631–1640.

[17]. Ryan J Gallagher, Kyle Reing, David Kale, and Greg Ver Steeg. 2017. Anchored correlation explanation: Topic modeling with minimal domain knowledge. *Transactions of the Association for Computational Linguistics* 5 (2017), 529–542.

[18]. Patrick Gebhard, Tobias Baur, Ionut Damian, Gregor Mehlmann, Johannes Wagner, and Elisabeth André. 2014. Exploring interaction strategies for virtual characters to induce stress in simulated job interviews. In *Proceedings of the 2014 international conference on Autonomous agents and multi-agent systems*. International Foundation for Autonomous Agents and Multiagent Systems, 661–668.

[19]. Thomas Gordon. 1977. *Leader Effectiveness Training*. New York: Wyden books. p. 57.

[20]. Sambhav Gupta, Krithika Jagannath, Nitin Aggarwal, Ramamurti Sridar, Shawn Wilde, and Yu Chen. 2019. Artificially Intelligent (AI) tutors in the classroom: A need assessment study of designing chatbots to sup-port student learning. In *Proceedings of the 2019 PACIS Pacific Asia Conference on Information Systems*. AIS.

[21]. Matt Grech. 2017. The Current State of Chatbots in 2017. (Apr 2017). https://getvoip.com/blog/2017/04/21/the- current- state- of- chatbots- in- 2017/

[22]. Jonathan Grudin and Richard Jacques. 2019. Chatbots, Humbots, and the Quest for Artificial General Intelligence. In *Proceedings of the 2019 CHI*

[23]. Tianran Hu, Anbang Xu, Zhe Liu, Quanzeng You, Yufan Guo, Vibha Sinha, Jiebo Luo, and Rama Akkiraju. 2018. Touch Your Heart: A Tone-aware Chatbot for Customer Care on Social Media. In *Proceedings of the 2018 CHI Conference on Human Factors in Computing Systems*. ACM, 415.

[24]. Zhiting Hu, Xuezhe Ma, Zhengzhong Liu, Eduard Hovy, and Eric Xing. 2016. Harnessing deep neural networks with logic rules. *arXiv preprint arXiv:1603.06318* (2016).

[25]. IBM Watson Assistant. 2019. Retrieved from https://www.ibm.com/cloud/watson-assistant/

[26]. Mohit Jain, Pratyush Kumar, Ishita Bhansali, Q Vera Liao, Khai Truong, and Shwetak Patel. 2018. FarmChat: A Conversational Agent to Answer Farmer Queries. *Proceedings of the ACM on Interactive, Mobile, Wearable and Ubiquitous Technologies* 2, 4 (2018), 170.

[27]. Douglas Jones, 1979, *Elementary Information Theory*. Clarendon Press.

[28]. Juji document for chatbot designers. 2019 Retrieved from https://docs.juji.io/

[29]. Soomin Kim, Joonhwan Lee, and Gahgene Gweon. 2019. Comparing Data from Chatbot and Web Surveys: Effects of Platform and Conversational Style on Survey Response Quality. In *Proceedings of the 2019 CHI Conference on Human Factors in Computing Systems*. ACM, 86.

[30]. Minha Lee, Sander Ackermans, Nena van As, Hanwen Chang, Enzo Lucas, and Wijnand IJsselsteijn. 2019. Caring for Vincent: A Chatbot for Self-Compassion. In *Proceedings of the 2019 CHI Conference on Human Factors in Computing Systems*. ACM, 702.

[31]. Terri Lee, Krithika Jagannath, Nitin Aggarwal, Ramamurti Sridar, Shawn Wilde, Timothy Hill, and Yu Chen. 2019b. Intelligent Career Advisers in Your Pocket? A Need Assessment Study of Chatbots for Student Career Advising. (2019).

[32]. Jingyi Li, Michelle X. Zhou, Huahai Yang, and Gloria Mark. 2017. Confiding in and listening to virtual agents. In *Proceedings of the 22nd International Conference on Intelligent User Interfaces-IUI*, Vol. 17.

[33]. Shixia Liu, Michelle X Zhou, Shimei Pan, Yangqiu Song, Weihong Qian, Weijia Cai, and Xiaoxiao Lian. 2012. Tiara: Interactive, topic-based visual text summarization and analysis. *ACM Transactions on Intelligent Systems and Technology (TIST)* 3, 2 (2012), 25.

[34]. Gale M Lucas, Jonathan Gratch, Aisha King, and Louis-Philippe Morency. 2014. It's only a computer: Virtual humans increase willingness to disclose. *Computers in Human Behavior* 37 (2014), 94–100.

[35]. Stephen Louw, R Watson Todd, and P Jimarkon. 2011. Active listening in qualitative research interviews. In *Proceedings of the International Conference: Research in Applied Linguistics, April*.

[36]. Ryan Thomas Lowe, Nissan Pow, Iulian Vlad Serban, Laurent Charlin, Chia-Wei Liu, and Joelle Pineau. 2017. Training end-to-end dialogue systems with the ubuntu dialogue corpus. *Dialogue & Discourse* 8, 1 (2017), 31–65.

[37]. Manychat. 2019 Retrieved from https://manychat.com/

[38]. Michael C McCord, J William Murdock, and Branimir K Boguraev. 2012. Deep parsing in Watson. *IBM Journal of research and development* 56, 3.4 (2012), 3–1.

[39]. Clifford Nass, Jonathan Steuer, and Ellen R Tauber. 1994. Computers are social actors. In *Proceedings of the SIGCHI conference on Human factors in computing systems*. ACM, 72–78.

[40]. Maximilian Nickel, Kevin Murphy, Volker Tresp, and Evgeniy Gabrilovich. 2015. A review of relational machine learning for knowledge graphs. *Proc. IEEE* 104, 1 (2015), 11–33.

[41]. Jay Nunamaker, Derrick Douglas, Elkins Aaron, Burgoon Judee, and Patton Mark.2011 Embodied conversational agent-based kiosk for automated interviewing. *Journal of Management Information Systems* 28.1: 17-48.

[42]. Ioannis Papaioannou, Christian Dondrup, Jekaterina Novikova, and Oliver Lemon. 2017. Hybrid chat and task dialogue for more engaging hri using reinforcement learning. In *2017 26th IEEE International Symposium on Robot and Human Interactive Communication (RO-MAN)*. IEEE, 593–598.

[43]. Geoffrey Keppel. 1991 *Design and analysis: A researcher's handbook.* Prentice-Hall, Inc.

[44]. Radim Rehrek and Petr Sojka. 2011. Gensim—statistical semantics in python. *statistical semantics; gensim;* Python; LDA; SVD (2011).

[45]. Carl R Rogers and Richard E Farson. 1984. Active listening. Organizational Psychology, 4th Ed. *Englewood Cliffs,* NJ (1984), 255–266.

[46]. Iulian V Serban, Alessandro Sordoni, Yoshua Bengio, Aaron Courville, and Joelle Pineau. 2016. Building end-to-end dialogue systems using generative hierarchical neural network models. In *Thirtieth AAAI Conference on Artificial Intelligence.*

[47]. Louis Shao, Stephan Gouws, Denny Britz, Anna Goldie, Brian Strope, and Ray Kurzweil. 2016. Generating long and diverse responses with neural conversation models.

[48]. Yangqiu Song, Shimei Pan, Shixia Liu, Michelle X Zhou, and Weihong Qian. 2009. Topic and keyword re-ranking for LDA-based topic modeling. In *Proceedings of the 18th ACM conference on Information and knowledge management*. ACM, 1757–1760.

[49]. S. Shyam Sundar and Jingyoung Kim 2019. Machine Heuristic: When We Trust Computers More than Humans with Our Personal Information. In *Proceedings of the 2019 CHI Conference on Human Factors in Computing Systems*. ACM, 538.

[50]. Mukund Sundararajan, Ankur Taly, and Qiqi Yan. 2017. Axiomatic attribution for deep networks. In *Proceedings of the 34th International Conference on Machine Learning-Volume 70*. JMLR. org, 3319–3328.

[51]. Ella Tallyn, Hector Fried, Rory Gianni, Amy Isard, and Chris Speed. 2018. The Ethnobot: Gathering Ethnographies in the Age of IoT. In *Proceedings of the 2018 CHI Conference on Human Factors in Computing Systems*. ACM, 604.

[52]. David Traum. 2017 *Computational Approaches to Dialogue*. The Routledge Handbook of Language and Dialogue, Chapter 9, Weigand, E. (Ed.), New York.

[53]. Debbe Thompson, Karen W Cullen, Maria J Redondo, and Barbara Anderson. 2016. Use of relational agents to improve family communication in type 1 diabetes: methods. *JMIR research protocols* 5, 3 (2016), e151.

[54]. James Vincent. 2016. Twitter taught Microsoft's friendly AI chatbot to be a racist asshole in less than a day. Retrieved from https://www.theverge.com/2016/3/24/11297050/tay-microsoft-chatbot-racist

[55]. Quan Wang, Zhendong Mao, Bin Wang, and Li Guo. 2017. Knowledge graph embedding: A survey of approaches and applications. *IEEE Transactions on Knowledge and Data Engineering* 29, 12 (2017), 2724–2743.

[56]. Yansen Wang, Chenyi Liu, Minlie Huang, and Liqiang Nie. 2018. Learning to Ask Questions in Open-domain Conversational Systems with Typed Decoders. In *Proceedings of the 56th Annual Meeting of the Association for Computational Linguistics (Volume 1: Long Papers)*. 2193–2203.

[57]. Guoyin Wang, Chunyuan Li, Wenlin Wang, Yizhe Zhang, Dinghan Shen, Xinyuan Zhang, Ricardo Henao, and Lawrence Carin. 2018a. Joint embedding of words and labels for text classification. *arXiv preprint arXiv:1805.04174* (2018).

[58]. Alice Watson, Timothy Bickmore, Abby Cange, Ambar Kulshreshtha, and Joseph Kvedar. 2012. An internet-based virtual coach to promote physical activity adherence in overweight adults: randomized controlled trial. *Journal of medical Internet research* 14, 1 (2012), e1.

[59]. Harry Weger Jr, Gina R Castle, and Melissa C Emmett. 2010. Active listening in peer interviews: The influence of message paraphrasing on perceptions of listening skill. *The Intl. Journal of Listening* 24, 1 (2010), 34–49.

[60]. Joseph Weizenbaum and others. 1966. ELIZA—a computer program for the study of natural language communication between man and machine. *Commun. ACM* 9, 1 (1966), 36–45.

[61]. Alex C Williams, Harmanpreet Kaur, Gloria Mark, Anne Loomis Thompson, Shamsi T Iqbal, and Jaime Teevan. 2018. Supporting workplace detachment and reattachment with conversational intelligence. In *Proceedings of the 2018 CHI Conference on Human Factors in Computing Systems*. ACM, 88.

[62]. Jason D Williams, Kavosh Asadi, and Geoffrey Zweig. 2017. Hybrid code networks: practical and efficient end-to-end dialog control with supervised and reinforcement learning. *Proceedings of the 55th Annual Meeting of the Association for Computational Linguistics (Volume 1: Long Papers)*. 2017

[63]. Ziang Xiao, Michelle X Zhou, and Wai-Tat Fu. 2019a. Who should be my teammates: Using a conversational agent to understand individuals and help teaming. In *Proceedings of the 24th International Conference on Intelligent User Interfaces*. ACM, 437–447.

[64]. Ziang Xiao, Michelle X Zhou, Q Vera Liao, Gloria Mark, Changyan Chi, Wenxi Chen, and Huahai Yang. 2019b. Tell Me About Yourself: Using an AI-Powered Chatbot to Conduct Conversational Surveys. *arXiv preprint arXiv:1905.10700 (2019)*.

[65]. Chen Xing, Wei Wu, Yu Wu, Jie Liu, Yalou Huang, Ming Zhou, and Wei-Ying Ma. 2017. Topic aware neural response generation. In *Thirty-First AAAI Conference on Artificial Intelligence*.

[66]. Zhou Yu, Alan W Black, and Alexander I Rudnicky. 2017. Learning conversational systems that interleave task and non-task content. *Proceedings of the 26th International Joint Conference on Artificial Intelligence*. AAAI Press, 2017.

[67]. Tiancheng Zhao, Ran Zhao, and Maxine Eskenazi. 2017. Learning Discourse-level Diversity for Neural Dialog Models using Conditional Variational Autoencoders. In *Proceedings of the 55th Annual Meeting of the Association for Computational Linguistics (Volume 1: Long Papers)*. 654–664.

[68]. Hao Zhou, Minlie Huang, Tianyang Zhang, Xiaoyan Zhu, and Bing Liu. 2018. Emotional chatting machine: Emotional conversation generation with internal and external memory. In *Thirty-Second AAAI Conference on Artificial Intelligence*.

[69]. Michelle X Zhou, Gloria Mark, Jingyi Li, and Huahai Yang. 2019. Trusting Virtual Agents: The Effect of Personality. *ACM Transactions on Interactive Intelligent Systems (TiiS)* 9, 2-3 (2019), 10.